\documentclass{aa}
\usepackage{graphicx}
\usepackage{txfonts}
\usepackage{amsmath}
\usepackage{amsfonts} 
\usepackage{lscape}
 
\def\kms{km~s$^{-1}$}
\def\teff{$\rm T_{\rm eff}$}
\def\gr{$\log {\rm g}$}
\def\vt{${\rm v_{\rm t}}$}
\def\ali{${\rm A(Li)_{NLTE}}$}
\def\he{HE\,0057-5959}
\def\lum{${\rm log(L/L_{\odot})}$}

\begin{document}

\title{The true nature of HE\,0057-5959, the most metal-poor Li-rich star}

\author{
A. Mucciarelli\inst{1,2} \and
P. Bonifacio\inst{3} \and
L. Monaco\inst{4} \and
M. Salaris\inst{5,6} \and 
M. Matteuzzi\inst{1,2}
}

\institute{
Dipartimento  di  Fisica  e  Astronomia  “Augusto  Righi”,  Alma  Mater Studiorum, Universit\`a  di Bologna, Via Gobetti 93/2, I-40129 Bologna, Italy
\and
INAF - Osservatorio di Astrofisica e Scienza dello Spazio di Bologna, Via Gobetti 93/3, I-40129 Bologna, Italy
\and
GEPI, Observatoire de Paris, Universit{\'e} PSL, CNRS, Place Jules
Janssen, 92195 Meudon, France
\and
Universidad Andres Bello, Facultad de Ciencias Exactas, Departamento de Ciencias Físicas - Instituto de Astrofísica, Autopista Concepcion-Talcahuano 7100, Talcahuano, Chile
\and
Astrophysics Research Institute, Liverpool John Moores University, 146 Brownlow Hill, Liverpool L3 5RF, United Kingdom
\and
INAF - Osservatorio Astronomico d{'}Abruzzo, via M. Maggini, 64100, Teramo, Italy
}

\authorrunning{A. Mucciarelli et al.}
\titlerunning{The Li-rich metal-poor star HE0057-5959}

\abstract{The Li-rich stars are a class of rare objects with a surface lithium abundance, A(Li), that exceeds 
that of other stars in the same evolutionary stage. The origin of these stars is still debated and valuable routes 
are the Cameron-Fowler mechanism, mass transfer process in a binary system or engulfment of rocky planets or 
brown dwarfs. Metal-poor ([Fe/H]$<$--1 dex) stars are only a small fraction of the entire population of Li-rich stars. 
We observed with MIKE at the Magellan Telescope the metal-poor ([Fe/H]=--3.95$\pm$0.11 dex) giant star \he , 
deriving \ali=+2.09$\pm$0.07 dex. Such Li abundance is significantly higher by about 1 dex than that of other 
stars in the same evolutionary stage. A previous analysis of the same target suggested that its high A(Li) 
reflects a still ongoing First Dredge-Up process. We revise the nature of \he\ by comparing its stellar parameters and A(Li) 
with appropriate stellar evolution models describing Li depletion due to the deepening of the convective envelope. 
This comparison rules out that \he\ is caught during its First Dredge-Up, being this latter already ended according to the 
parameters of this star. Its A(Li), remarkably higher than the typical lithium plateau drawn by similar giant stars, demonstrates 
that \he\ joins the class of the rare metal-poor Li-rich stars.  \he\ is the most metal-poor Li-rich star discovered so far. 
We consider different scenarios to explain this star also comparing it with the other metal-poor Li-rich stars. 
No internal mixing able to activate the Cameron-Fowler mechanism is known for metal-poor stars at this evolutionary stage. 
Also the engulfment of planets is disfavoured because such metal-poor stars should not host planets. 
Finally, \he\ is one of the most Na-rich among the Li-rich stars and we found that 
a strong excess of Na abundance is common to all the three Li-rich stars with [Fe/H]$<$-3 dex. 
This finding could support the scenario of mass transfer from a massive companion star 
(able to simultaneously produce large amounts of both elements) in a binary system, even if we found no evidence of radial velocity variations.}


\maketitle

\section{Introduction}

Lithium is one of the few nuclei produced during the Big Bang nucleosynthesis.
Because of its fragility, the Li nuclei are immediately destroyed in stellar layers exceeding 
temperatures of $\sim2.5\cdot10^{6}$ K. 
In a low-mass star, the surface lithium abundance A(Li) changes with the time and its changes
reflect the main episodes of mixing occurring during the evolution of the star. 
When the convective envelope deepens into the stellar interior reaching stellar layers hot enough 
to burn Li, Li-free material is brought to the surface reducing A(Li).
The evolution of A(Li) with the luminosity (or the gravity) can be described as two phases with constant 
A(Li), namely the Spite Plateau \citep{spite82,bonifacio97,aoki09} and the lower red giant branch (RGB) Plateau 
\citep{mucciarelli12,mucciarelli22}, both followed by an abrupt drop corresponding to the First Dredge-Up
(FDU) and the RGB bump (RGBb) mixing episode, respectively. The A(Li) depletion at the FDU is a natural consequence of the standard stellar 
evolution \citep{iben67} while to explain the second drop at the RGBb non-canonical mixing 
processes have to be included, the most popular one being the thermohaline mixing \citep[see e.g.][]{charbonell20}.

Within this framework, the so-called Li-rich stars are a class of peculiar and rare objects whose A(Li)
significantly exceeds that measured in stars of similar evolutionary stage. In some cases \citep[see e.g.][]{koch11,kow22}, 
metal-poor stars have A(Li) higher than the abundance formed during the Big Bang and inferred by the baryon density obtained 
from Wilkinson Microwave Anisotropy Probe (WMAP) and Planck satellites \citep[A(Li)=+2.72$\pm$0.04 dex,][]{coc17},  while some metal-rich stars have A(Li) 
higher than that of the interstellar medium \citep[A(Li)$\sim$+3.3 dex,][]{asplund09}.
These very high A(Li) values  suggest that lithium is not preserved but other mechanisms of production should be at work.
Their frequency is $\sim$1 \% or less of all the stars \citep{kirby16,casey16,gao19,deepak20}. 
They have been observed at any metallicity \citep{ruchti11,martell13,gao19,sitnova23}, in any evolutionary stages \citep{kirby16,li18}, 
in the Milky Way halo \citep{ruchti11,li18}, thick disc \citep{monaco11}, thin disc \citep{casey16,deepak20} and bulge \citep{gonzalez09}, 
and dwarf spheroidal galaxies \citep{kirby12}, as well as 
in globular clusters \citep[GCs,][]{smith99,ruchti11,koch11,monaco12,dorazi15,kirby16,mucciarelli19,sanna20} 
and in open clusters \citep{monaco14}.

Li-rich stars could be explained invoking different classes of processes, namely internal production or 
external origin of the extra Li.
Three main groups of processes have been proposed to explain the existence of these stars, 
namely the Cameron-Fowler mechanism, mass transfer in binary systems and 
engulfment of small bodies.

(a) \citet{cameron55} and \citet{cf71} proposed a mechanism to produce fresh Li in 
Asymptotic Giant Branch (AGB) stars experiencing the hot bottom burning.
$^{7}$Li is produced after the decay of $^{7}$Be. 
However the temperatures needed to produce $^{7}$Be from $\alpha$-capture on $^{3}$He
are one order of magnitude higher than the temperature of Li burning. Therefore, 
new Li is immediately destroyed but if $^{7}$Be is fast transported toward cooler regions,  Li can survive.
The Cameron-Fowler mechanism, originally proposed for intermediate-mass 
AGB stars \citep[indicatively in the mass range 4-8 $M_{\odot}$, see e.g.][]{sack92,ventura11}, 
can work whenever a mechanism carries $^{7}$Be to cooler stellar regions. 
In the case of AGB stars, this process is driven by the convection, being 
the bottom of the convective envelope hot enough to produce $^{7}$Be. 
The existence of Li-rich stars at different metallicity, mass and evolutionary stage 
can be explained with the Cameron-Fowler mechanism only invoking extra mixing, i.e. 
thermohaline mixing \citep{charbonell05}, magneto-thermohaline \citep{denissenkov09}, 
mass-loss mechanisms occurring in RGB \citep{delareza96}.
An additional mechanism able to trigger the Cameron-Fowler mechanism is the ingestion of 
sub-stellar companions, like rocky planets or brown dwarfs causing 
an increase of angular momentum and a rotationally induced mixing \citep{denissenkov04}.

(b) Accretion of matter from a companion with an enhancement of Li, for instance any star 
able to trigger the Cameron-Fowler mechanism. Intermediate-mass AGB stars are the natural candidates, 
assuming that the low-mass Li-rich star belongs to a binary system with a massive companion 
(now evolved as a white dwarf). Possible additional signatures of this process could be 
an enhancement of those elements produced in the interiors of AGB stars (i.e., carbon and slow 
neutron-capture elements) and radial velocity variability.

(c) engulfment of sub-stellar companions, like rocky planets, hot Jupiters, brown dwarfs, 
having high Li enhancement \citep{siess99}. This process should be coupled with infrared excess, 
possible strong magnetic fields and X-ray activities.

In this paper, we discuss the metal-poor, giant star \he\ (Gaia EDR3 4903905598859396480) 
whose A(Li) has been previously derived by \citet{jacobson15} and 
interpreted as a normal giant star caught during its still ongoing FDU.

\section{MIKE Observations and chemical analysis}
\label{par}

\subsection{Observations and radial velocity}
The target star \he\ was observed with the Magellan Ianmori Kyocera Echelle 
(MIKE) spectrograph \citep{bernstein03} mounted at the Magellan II Telescope at Las Campanas Observatory, 
under the program CN2017B-54 (PI: Monaco) during the night 2017 October 6$^{th}$.
We adopted a 0.7''X5.0'' slit corresponding to spectral resolution of 53,000 and 42,000, 
in the blue and red arm respectively and covering the spectral range between $\sim$3400 \AA\ and $\sim$9400 \AA . 
A total exposure time of 3 by 3000 sec was adopted, providing a signal-to-noise ratio for pixel of 80 around the 
Li line at 6708 \AA .
The spectral reduction, including bias-subtraction, flat-fielding, spectral extraction and wavelength 
calibration was performed  with the dedicated CarPy pipeline \citep{kelson03}.
The heliocentric radial velocity (RV) was measured through a cross-correlation against a synthetic spectrum 
as template, obtaining +376.2$\pm$0.2 \kms\ , in agreement with the previous estimates 
by \citet[][+375.3 \kms ]{norris13}, \citet[][+376.7 \kms ]{jacobson15}  
and \citet[][+377.90$\pm$3.98 and +378.23$\pm$1.47 \kms ]{arentsen19}. 
No RV values are provided by Gaia. 

\subsection{Atmospheric parameters}
\label{atmpar}

The effective temperature (\teff ) and the surface gravity (\gr ) were estimated by the photometry 
in order to avoid significant biases affecting the spectroscopic determinations of these parameters in metal-poor giant stars 
\citep[see][]{mb20}. We adopted magnitudes from the early third data release of the ESA/$Gaia$ mission 
\citep{prusti16,brown20} and a colour excess of  E(B-V)=~0.017 mag by \citet{schlafly11}. 
The extinction coefficients of the three Gaia bands were derived adopting the iterative procedure described in \citet{lombardo21}.
Effective temperature was estimated using the $(G_{BP}-G_{RP})_0$-\teff\  transformation by \citet{mbm21}, 
deriving \teff = 5420$\pm$80 K, where the uncertainty includes those arising from the photometric color, 
the colour excess and the adopted colour-\teff\ transformation.
 We check that \teff\ derived from the other two Gaia colours, $(G_{BP}-G)_0$ and $(G-G_{RP})_0$, are 
in excellent agreement with that by $(G_{BP}-G_{RP})_0$, \teff =5477 and 5426 K, respectively.

Stellar luminosity (${\rm log(L/L_{\odot})}=-0.4\cdot({\rm M_{bol}}-{\rm M_{\odot})}$) was calculated 
adopting the photogeometric distance posterior from the Gaia parallax provided by \citet{bj21} and the bolometric correction
calculated from a new grid of synthetic fluxes (Mucciarelli et al. in prep.) computed with the code {\tt ATLAS9} \citep{Kurucz2005}.  
Surface gravity (\gr = --10.32 + log(M) +4$\cdot$\teff - log(L) ) was obtained adopting the photometric \teff\ and  
the stellar luminosity described above and a stellar mass equal to 0.76 ${\rm M_{\odot}}$ that is a reasonable value 
for RGB stars with old ages and low metallicities, according to theoretical isochrones by \citet{pietr21}. 
Ages in the range of $\sim$11-13 Gyr, reasonable for a very metal-poor star like \he\ , provide very similar stellar 
masses along the RGB, lower than $\sim$0.82 ${\rm M_{\odot}}$ with a negligible impact (less than 0.04 dex) on \gr.
In the quoted \gr\ uncertainty we accounted for uncertainties in \teff\ , luminosity and adopted stellar mass.

Microturbulent velocity was estimated spectroscopically by minimising any trend between the abundances 
from Fe~I lines and their reduced equivalent widths. 
The final parameters for \he\ are \teff\ =~5420$\pm$80 K, \gr\ =~3.05$\pm$0.10, \lum\ = 1.18$\pm$0.14, and \vt =~1.5$\pm$0.2 \kms  
see Table ~\ref{tabnot}. 
These values well agree with those estimated by \citet{jacobson15},   \teff\ =~5413 K, \gr\ =~3.41, and \vt\ =1.4 \kms .
On the other hand, the temperature provided by \citet{norris13} is cooler (5257 K) and derived 
as the average of temperatures from spectrophotometric flux and Balmer lines. For the chemical analysis 
of this star, \citet{yong13} adopted \teff\ by \citet{norris13}, while a new value of \gr\ (2.65 dex)  was derived 
based on the stellar temperature and a theoretical isochrone with appropriate metallicity and age of 10 Gyr. 
Their lower \gr\ reflects mainly the 
difference in \teff\ between our and their analysis.

\begin{table}
\caption{Stellar parameters for the target star \he\ .}       
\label{tabnot}     
\centering                          %
\begin{tabular}{c  c }       
\hline\hline                 
{\rm Parameter} &  {\rm Value}  \\
\hline
\hline                        
\teff\  &    5420$\pm$80 K \\
\gr\    &    3.05$\pm$0.10 \\
\vt\    &    1.5$\pm$0.2 \kms \\ 
\lum\   &   1.18$\pm$0.14  \\

\hline                                   
\end{tabular}
\end{table}

\subsection{Chemical analysis}
\label{cheman}

\begin{figure}
\centering
\includegraphics[scale=0.7]{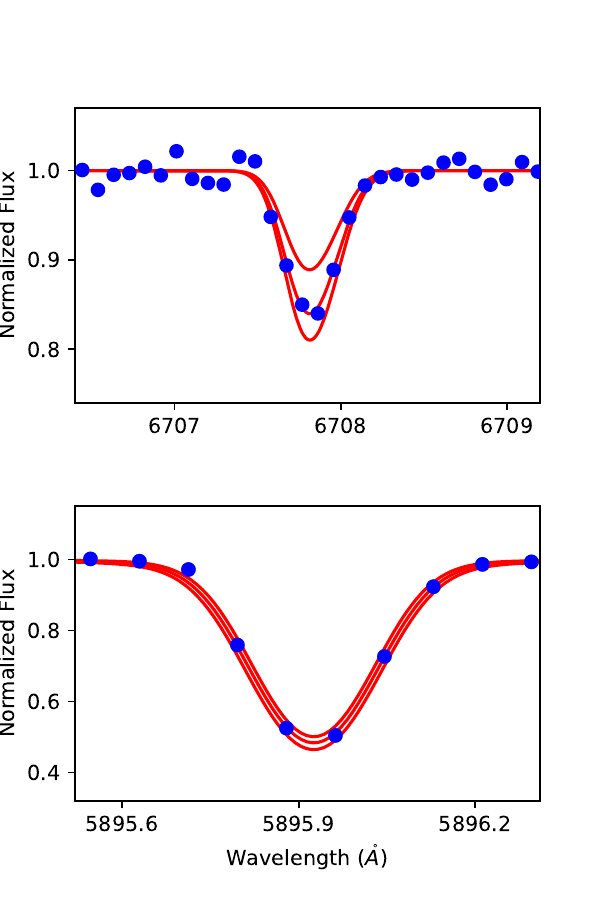}
\caption{
Spectral regions of the MIKE spectrum  (blue points) around the 
Li resonance line and a Na D line (upper and lower 
panel, respectively), with superimposed synthetic spectra calculated with the best-fit abundance 
(central red curve) and $\pm$0.1 dex from the best-fit abundance.} 
\label{ex_fit}
\end{figure}

Abundances of Mg, Al, Si, Ca, Ti and Fe were obtained from measured equivalent widths (EWs) using the code 
{\tt GALA} \citep{gala}, while for atomic transitions affected by blending and/or isotopic/hyperfine splittings 
(Li, Na, Sr and Ba) and  molecular features (G-band for C), the abundances were derived 
through a $\chi^{2}$-minimisation between the observed lines  and grids of synthetic spectra 
calculated with {\tt SYNTHE} \citep{Kurucz2005}. 
For other species, spectral lines are too weak and undetectable because of the very low metallicity 
of the target star, coupled with its relatively high \teff\ , preventing also useful upper limits. 
We checked also the possibility to measure N and $^{12}$C/$^{13}$C but we cannot identify $^{13}$C transitions 
or CN molecular bands at 3870-3885 \AA\ . Despite the importance of N and $^{12}$C/$^{13}$C, the derived 
upper limits are not meaningful ([N/Fe]$<$+3.5 dex, $^{12}$C/$^{13}$C$>$5) and they do not provide useful insights 
for the interpretation of the target star.
In this analysis, we adopted an {\tt ATLAS9} model atmosphere \citep{Kurucz1993,Kurucz2005} 
calculated with a new opacity distribution function (Mucciarelli et al. in prep.) with [Fe/H]=--4.0 dex 
and [$\alpha$/Fe]=+0.4 dex.\\
Abundance ratios scaled on the solar abundances by \citet{lodders10} and \citet{caffau11} are 
listed in Table~\ref{resabu}, together with the number of measured lines and the total uncertainty. 
The latter was computed by adding in quadrature the internal error and that arising from the adopted 
atmospheric parameters (uncertainties in the derived parameters are discussed in Section~\ref{atmpar}). 
The internal error is estimated as the standard error of the mean ($\sigma/\sqrt({\rm N_{lines}})$) 
when at least two lines are measured, while for element with one only transition we consider 
the uncertainty on the abundance as obtained from the EW error or from Montecarlo simulations, 
for lines measured from EWs and spectral synthesis respectively \citep[see details in][]{mucciarelli13b}.
We report in Table~\ref{resabu} both LTE and NLTE abundances, in the latter case adopting the corrections 
by \citet{wang21} for Li, by \citet{lind11} for Na, by \citet{nordlander17} for Al, by \citet{bergemann13} 
for Si and by \citet{mashonkina23} for the other elements. 
The atomic data of the measured atomic lines are listed in Appendix~\ref{llist}. \\
We derived [Fe/H]=--3.95$\pm$0.11 dex. Departures from LTE increase the Fe abundance of this star by $\sim$+0.22 dex 
\citep{mashonkina23}. In the following discussion we refer to the LTE [Fe/H] for consistency with the [Fe/H] values 
derived for all the other Li-rich metal-poor stars known so far but the use of NLTE [Fe/H] does not change 
our conclusion about the nature of \he\ .
The star is enhanced in [$\alpha$/Fe] (Mg, Si, Ca and Ti) and characterised by sub-solar [Sr/Fe] and [Ba/Fe]. 
Also, the star is enhanced in [C/Fe] (+1.07 dex), indicating that it can be labelled as CEMP-no (enhanced in C but 
not in s-process elements), as already suggested by \citet{jacobson15}.
The comparison with the abundances measured by \citet{yong13} and \citet{jacobson15} is discussed 
in Appendix~\ref{compar}.\\
Lithium abundance was derived from the resonance line at 6708 \AA\ , while the subordinated Li 
line at 6103 \AA\ is not detected. 
The 3D-NLTE correction by \citet{wang21} was applied to the Li abundance, leading to a final abundance 
of \ali=+2.09$\pm$0.07, consistent with the value quoted by \citet{jacobson15}, \ali=+1.97 dex. 
The NLTE correction for this star is of +0.04 dex.\\
Sodium abundance is derived from the Na D lines at 5890 and 5896 \AA\ that are the only Na transitions 
visible in the spectrum because of the very low metallicity of the star (also \citet{yong13} and \citet{jacobson15} 
derived the Na abundance of \he\ measuring these two lines only). 
We checked that the Na D lines are not contaminated by Na interstellar lines because of the high RV of the star (see Figure~\ref{naspec}).
The NLTE corrections for Na abundance by \citet{lind11} were applied, leading to [Na/Fe]=+1.92$\pm$0.08 dex,
in agreement with the values listed by \citet{yong13} and \citet{jacobson15}. 
The Na doublet at 5682-88 \AA\ is too weak to be observed and we can only derive an upper limit, [Na/Fe]$<$2.1 dex.
Fig.~\ref{ex_fit} shows as examples the best-fit synthetic spectra obtained for Li and Na lines.

\begin{table*}
\caption{LTE and NLTE chemical abundances measured in \he\ , together with the number of used lines 
and the total uncertainty. The abundances are reported as [X/Fe] but for Li (A(Li)) and Fe ([Fe/H]).
The NLTE abundance ratios [X/Fe] are calculated assuming the NLTE [Fe/H].}            
\label{resabu}     
\centering                          %
\begin{tabular}{c  c c  c c}       
\hline\hline                 
Ion &   $N_{lines}$ & {\rm LTE} & {\rm NLTE} & $\sigma$  \\
\hline
 &  & (dex) & (dex) &  (dex)  \\
\hline                        
 {\rm A(Li)}     &  1   &   +2.05   &  2.09 &  0.07 \\
 {\rm [C/Fe]}    &  1   &   +1.07   &   ---   & 0.10  \\
 {\rm [Na/Fe]}   &  2   &   +1.92   & +1.17   &  0.08 \\
 {\rm [Mg/Fe]}   &  4   &   +0.54   & +0.41   & 0.05  \\
 {\rm [Al/Fe]}   &  2   &  --0.11   & +0.29   & 0.16 \\
 {\rm [Si/Fe]}   &  1   &   +0.85   & +0.63   &  0.15  \\
 {\rm [Ca/Fe]}   &  1   &   +0.44   & +0.22   &   0.07\\
 {\rm [Ti~II/Fe]}&  3   &   +0.44   & +0.41   &  0.10 \\
 {\rm [Fe/H]}  &  21  &  --3.95   & --3.75   & 0.11  \\
 {\rm [Sr~II/Fe]}&   1  &  --1.11   & --1.12   & 0.15 \\
 {\rm [Ba~II/Fe]}&   1  &  --1.47   & --1.49   & 0.20  \\
\hline                                   
\end{tabular}
\end{table*}

\section{The evolutionary stage of HE0057-5959}

\begin{figure*}
\centering
\includegraphics[scale=0.5]{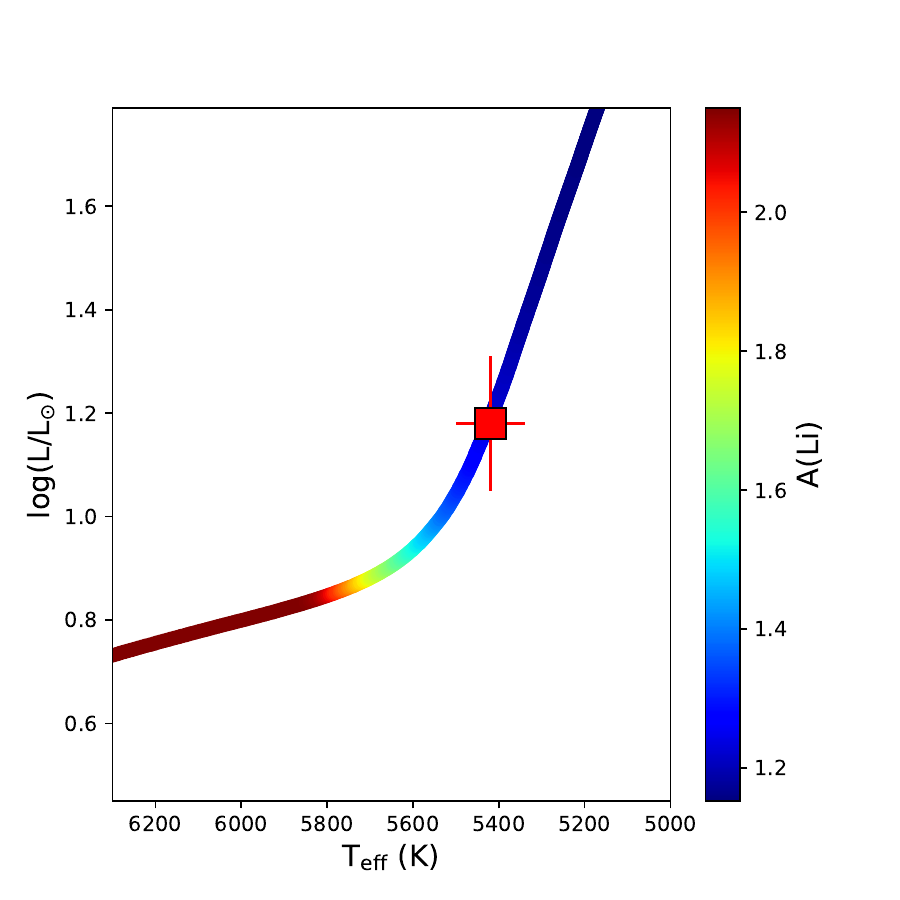}
\includegraphics[scale=0.65]{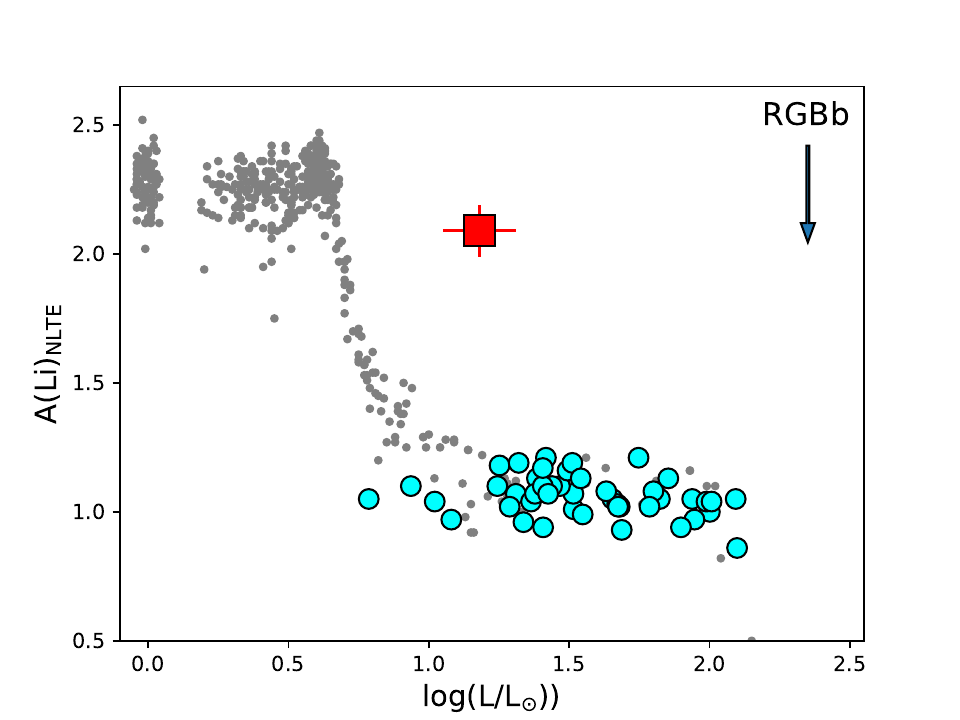}
\caption{
Left panel: position of the target star \he\ in the Hertzsprung-Russell diagram (red square) in comparison with 
the theoretical track of a star with 0.76 $M_{\odot}$, [Fe/H]=--4.0 dex and [$\alpha$/Fe]=+0.4 dex, 
coloured according to the predicted A(Li). The colour scale is shown on the right side.
The model is re-scaled in order to match the average \ali\ of the RGB stars by \citet{mucciarelli22}.
Right panel: \ali\ as a function of \lum\ for \he\ (red square) 
in comparison with the RGB stars sample by \citet[][cyan circles]{mucciarelli22} 
and the stars of the metal-poor ([Fe/H]$\sim$--2.0 dex) globular cluster NGC~6397 \citep[][grey circles]{lind09}. 
 The luminosity level of the RGBb for the metallicity of \he\ is marked with an arrow.}

\label{lit1}
\end{figure*}

\citet{jacobson15} derived \ali\ for a sample of 24 metal-poor giant stars, including \he . 
All these stars are fainter than that of the RGBb (with \teff\ between $\sim$4700 and $\sim$5400 K) 
and therefore they are not affected by the extra-mixing episode associated to the RGBb. 
The sample stars have very similar \ali , around +1.0 dex, well matching the values observed by \citet{mucciarelli22} 
for stars in the same evolutionary stage. In the \citet{jacobson15} sample \he\ is a clear outlier, with 
a significantly higher lithium abundance (\ali=+1.97 dex) 
despite having only a slightly higher \teff\ and lower \lum\ than the other sample stars. 
\citet{jacobson15} quoted for this star \teff\ = 5413 K, while the hottest star in the remaining sample 
has \teff\ = 5260 K and \ali\ = 1.21 dex. They do not recognise \he\ as a Li-rich star, claiming that 
this star has a A(Li) depletion level appropriate to its \teff\ . In other words, they interpret this star as a Li-normal star 
with an high \ali\ due to a still ongoing FDU, the only mechanism able to justify the measured \ali\ without invoking 
an anomalous enhancement of lithium. \\
Stellar parameters and \ali\ from our new analysis agree with those by \citet{norris13} and \citet{jacobson15} and 
we confirm that \he\ belongs to the first ascent of the RGB (see left panel of Fig.~\ref{lit1}). 
The luminosity of the target is fainter than the RGBb (occurring at \lum$\sim$2.35 for the metallicity of the star, 
see left panel of Fig.~\ref{lit1}.) 
confirming that \he\ is not affected by the extra-mixing episode associated to the RGBb.
In order to establish whether its \ali\ is compatible with an ongoing FDU, 
we compared the \ali\ of \he\ with that predicted by a theoretical model 
\citep[calculated with the same code and input physics as in][]{pietr21}
of a star with 0.76 $M_{\odot}$ and Z=3.3$\cdot10^{-6}$ (corresponding to [Fe/H]=--4.0 dex 
and [$\alpha$/Fe]=+0.4 dex) and without atomic diffusion.\\ 
As clearly visible in left panel of Fig.~\ref{lit1}, at this metallicity the drop of A(Li) due to the FDU occurs at the end of the 
sub-giant branch  and it is confined in a narrow region in \teff , 5500--5800 K, and \lum , 0.8--1.0.
The stellar parameters of \he , within their uncertainties, are not compatible with the position of the FDU 
and they indicate that the star has already finished the FDU. The same conclusion is reached whether 
the stellar parameters derived by \citet{norris13} and \citet{jacobson15} are adopted, see Section~\ref{atmpar}.
If we assume that this star before the FDU had \ali\ similar to that of the Spite Plateau stars ($\sim$2.2-2.3 dex), 
the measured abundance (\ali\ =+2.09 dex) should be observed when the star starts its FDU, 
at \teff\ hotter by $\sim$300-400 K and  \lum\ fainter by 0.4 dex with respect to those of \he .
Therefore, the measured \ali\ of \he\ is not compatible 
with what we expect for a star in that evolutionary stage starting from the Spite Plateau. 
Also, \ali\ of \he\ is clearly incompatible with the abundances of lower RGB stars that share a very similar A(Li) close to +1.0 dex 
(see right panel of Fig.~\ref{lit1}).
Hence, we revise the nature of \he , claiming that it is a genuine Li-rich star, having 
a lithium abundance significantly higher than those measured in stars at the same evolutionary stage 
and not compatible with a still ongoing FDU.
This is the most metal-poor Li-rich star discovered so far, followed by the giant star LAMOST J070542.30+255226.6 
with [Fe/H]=--3.12 dex \citep{li18}.

\section{Metal-poor Li-rich stars in the literature}
\label{literature}

\begin{figure}[h]
\centering
\includegraphics[scale=0.48]{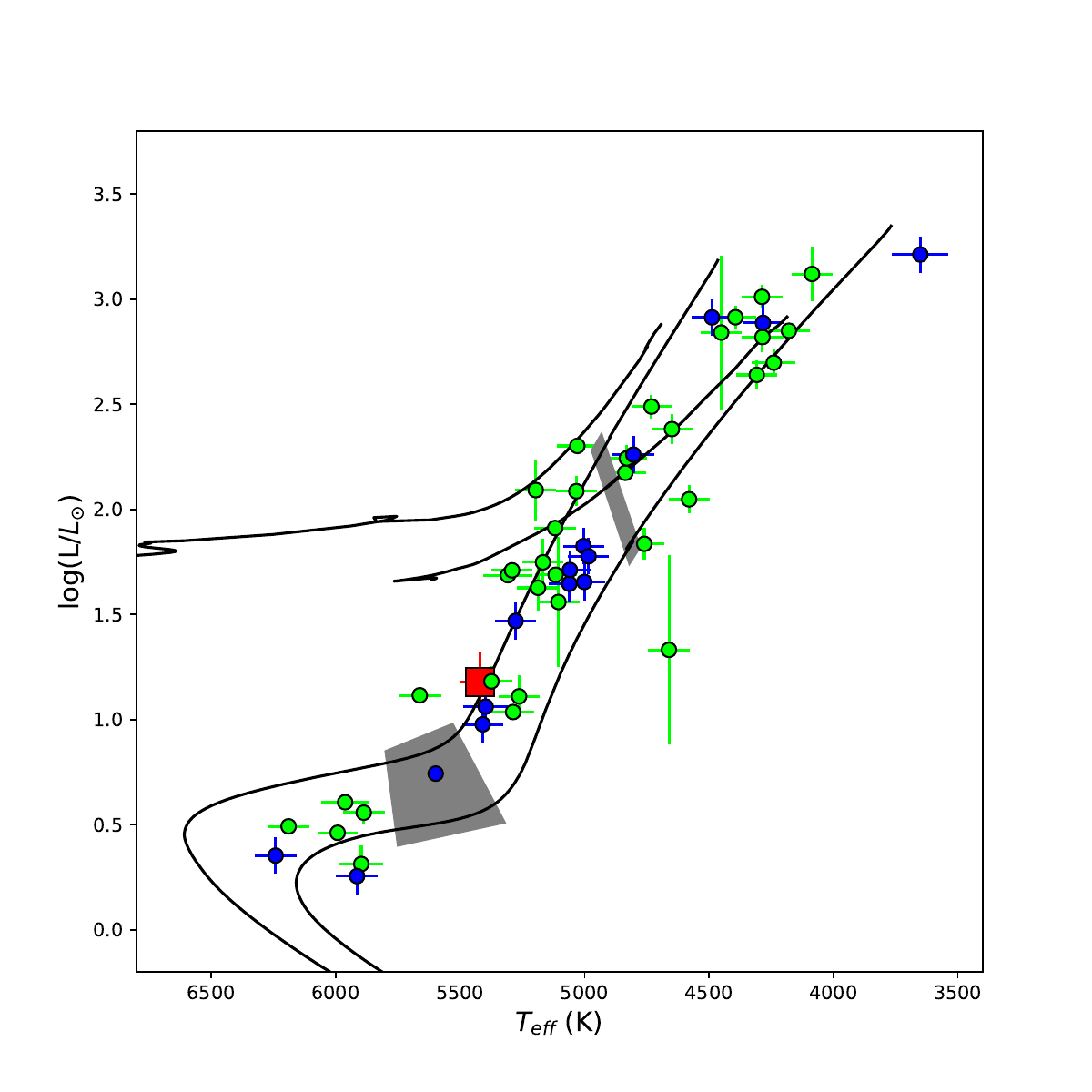}
\caption{Position in the Hertzsprung-Russell diagram of the target star \he\ (red square) 
and of the other metal-poor ([Fe/H]$<$--1 dex) Li-rich stars discovered so far (see details in Section~\ref{literature}): 
green circles are Galactic field stars 
and blue circles are globular clusters stars.
Two theoretical isochrones with [Fe/H]=--3.2 and --1.2 dex, 
$\alpha$-enhanced chemical mixture and an age of 13 Gyr \citep{pietr21} are shown as reference (black curves).
The two grey shaded areas indicate the loci 
where the drops of \ali\ due to the FDU and further extra-mixing episodes, respectively, occur.}
\label{lit2}
\end{figure}

In order to properly compare the properties of \he\ with those of other Li-rich stars, 
we collected a database of all the metal-poor ([Fe/H]$<$--1.0 dex) Li-rich stars discovered so far 
in our Galaxy, performing an effort to homogenise their stellar parameters and \ali .
Adopting for each star the literature stellar parameters and LTE A(Li) derived from the 6708 \AA\ line, we calculated
the corresponding synthetic profile and its equivalent width by integration. After that, 
we calculated the curve of growth for the Li line assuming the new stellar parameters and we derived 
the new A(Li) based on the previous equivalent width. The average difference between the LTE A(Li) values 
obtained with the new and literature parameters are +0.01$\pm$0.03 dex ($\sigma$=0.18 dex) with only three stars 
with an absolute discrepancy larger than 0.3 dex. Excluding these three stars the average difference 
is +0.04$\pm$0.02 dex ($\sigma$=0.12 dex).
For all the stars we applied the 3D-NLTE corrections by \citet{wang21}. 
Table~\ref{tab_field} and ~\ref{tab_gc} list for all the targets the new \teff , \gr , \lum , \ali\ and 
the literature values of [Fe/H] and [Na/Fe].
The position of all the metal-poor Li-rich stars in the Hertzsprung-Russell diagram is shown in Fig.~\ref{lit2} 
where we highlight the mean loci of the FDU and RGBb. We show also two theoretical isochrones 
with age of 13 Gyr and [Fe/H]=--3.2 and --1.2 dex \citep{pietr21} as a reference to identify the evolutionary stage of 
the Li-rich stars.

\subsection{Galactic field Li-rich stars}
Aside \he , 34 metal-poor Li-rich stars (all of them with \ali\ higher than the average A(Li) measured in stars 
with the same evolutionary stage) have been discovered in the Milky Way field, 
covering a range of metallicity between [Fe/H]$\sim$--3.1 and $\sim$--1.4 dex. 
All these stars were recovered in the Gaia eDR3 catalogue and 
their parameters (\teff , \gr\ and \lum ) were derived using the same approach described in Section~\ref{par}. 
We adopted the colour excess from the latest 3-dimensional dust extinction maps by \citet{lallement22} for the closest targets, 
and values from \citet{schlafly11} for those targets whose distance is outside the maps by \citet{lallement22}. 
The uncertainties in \teff\ are dominated by the precision of the adopted colour-\teff\ transformation 
\citep[$\sim$80 K, see][]{mbm21}, with a negligible contribution arising from photometric error and reddening. 
The uncertainties in \lum\ are smaller than 0.1 dex and dominated by the distance error. Only for four stars 
(namely SDSS J093627.44+293535.7, 2MASS J10122548-2030068, 2MASS J16070923+0447126 and SDSS J143207.14+081406.1) errors in \lum\ 
are higher than 0.2 dex, reflecting their large uncertainty in the Gaia parallaxes.  
For seven stars with \gr\ lower than 1.3 \citep[the boundary of the grid by][]{wang21}
we adopted \gr=1.3 to calculate the 3D-NLTE correction.
The new parameters well agree with the literature ones, with average differences (this study - literature) 
of +19$\pm$23 K ($\sigma$=137 K) for \teff\ and +0.02$\pm$0.03 ($\sigma$=0.20) for \gr . 
The largest differences are in \teff\ for the stars LAMOST J055408.54+523559.0, LAMOST J075816.39+470343.3 \citep{li18}, 
2MASS J05241392-0336543 \citep{kow22} and  UCAC4 212-183136 \citep{susmitha24}, 
with differences of --330, --254, --300 K and +228 K, respectively.

\subsection{Galactic globular clusters Li-rich stars}
16 Li-rich stars have been discovered so far in Galactic GCs. 
We recovered 15 of them in the Gaia catalogue, excluding the cepheid star V42 in M5 \citep{carney98} 
because its variability leads to large uncertainties in the stellar parameters.
Three Li-rich stars have a double identification: Stet-M68-S232 in M68 \citep{ruchti11,kirby16}, 
M3-IV 101 in M3 \citep{kraft99,ruchti11} and 132 in M30 \citep{kirby16,gruyters16}.
Their stellar parameters were derived adopting [Fe/H], distance and E(B-V) by \citet{harris} 
but for the two Li-rich stars in $\omega$ Centauri \citep{mucciarelli19} for which we adopted their proper [Fe/H] values.
For the Li-rich stars in NGC~1261 and NGC~6397 we assumed A(Li)=4.0 dex to avoid extrapolation 
in the 3D-NLTE grids. For other three stars (namely M3-IV in M3, Stet-M68-S232 in M68 and V2 in NGC~362) 
we assumed \gr =~1.3 as explained above for the field Li-rich stars. \\
The Gaia \teff\ well agree with the values quoted in the literature, with an average difference 
of +11$\pm$30 K ($\sigma$=116 K) and no relevant outliers but the cold star V2 in NGC~362 \citep{smith99} 
that we found 250 K cooler.
Concerning \gr\ the average difference is +0.06$\pm$0.04 ($\sigma$=0.17); most of the targets have 
differences in \gr\ within $\pm$0.1 with three main exceptions: V2 in NGC~362 that has \gr\ lower by 0.4 dex 
than the \citet{smith99} value,  97812 in NGC~3201 \citep{agu22} and M3-IV in M3 \citep{ruchti11} that 
have \gr\ higher than the literature by 0.4 dex.

\section{Discussion}

\subsection{The Li-rich star \he\ }
The star \he\ discussed in this work is the most metal-poor Li-rich star identified so far. 
What is the origin of its Li enhancement? We consider the main mechanisms previously proposed 
to explain Li-rich stars.

Concerning the scenario of internal Li production, 
the main issue for this star is the lack of an internal mixing process able to activate the 
Cameron-Fowler mechanism. Due to the low metallicity of the star, its shallow convective envelope 
cannot reach the layers where $^{7}$Be is produced through $\alpha$-captures on $^{3}$He and this region 
is not in contact with the surface, preventing any internal mixing.
Possible extra-mixing mechanisms can occur in low-mass stars but at the RGBb \citep{charbonell00,palacios01} 
or close to the He-flash \citep{silva14}, while this star is located at the base of the RGB.

Mass transfer process from a more massive, now evolved, companion star able to produce fresh Li 
from the Cameron-Fowler mechanism is a valuable and simple route to explain Li-rich stars. 
The measured RV from our MIKE spectrum nicely matches the previous estimates 
indicating no large RV variations. 
The RUWE value of the star is 0.98, indicating a well-behaved single star astrometric solution. 
However, wide binary systems can have too long periods to be identified with the Gaia observations. 
We cannot totally rule out that the star belongs to a long-period or highly inclined binary.
In this case, the enhancement of Li could be explained as the result 
of a mass transfer process from a more massive companion star during its AGB stage. 
We note that none of the Li-rich stars for which multiple epochs are available 
has shown evidence of variability, with the only exception of 25664 in $\omega$ Centauri 
\citep{mucciarelli21}.

The Li overabundance in a giant star can be the result of engulfment of a planet as the star
evolves on the RGB, increasing its radius. This seems to be the most likely explanation for
the solar metallicity Li-rich giant BD+48\,740 \citep{adamov12}.
However the extremely low metallicity of \he\ makes it unlikely that it hosts, or has hosted planets. 
In spite of early claims that the frequency of planets around giant stars is not correlated to metallicity
\citep{pasquini07}, subsequent investigations found that the higher the metallicity,
the higher the probability of hosting a planet \citep[][and references therein]{wolthoff22}, 
similar to what found for dwarf stars \citep[][and references therein]{adibekyan19}.
To our knowledge the two planet-host stars with the lowest metallicity are BD+20\,2457 \citep{maldonado13} and
24 Boo \citep{takarada18}, both of metallicity around --0.8 dex.
Clearly if we were to extrapolate this metallicity dependence of hosting planets down to --4.0 dex
we would found a very small number, practically zero.
One should however keep in mind that the stars with metallicity below --3.0 dex
are not represented in any of the planet search surveys, thus such an extrapolation
cannot be supported by any data.
We nevertheless believe that, based on our current knowledge, the possibility
that the high Li abundance in \he\ is due to planet engulfment can be discarded.

In conclusion, the mass transfer scenario remains the most promising one, lacking a well established mechanism 
able to induce a Cameron-Fowler mechanism in a low-mass, very metal-poor star like \he . 
Whatever mechanism is able of generating a Li-rich star, that mechanism must also occur down to [Fe/H]$\sim$--4 dex.

\subsection{An overview on the metal-poor Li-rich stars}

We discuss the properties of \he\ in comparison with the other metal-poor ([Fe/H]$<$--1 dex) Li-rich stars 
(see Section~\ref{literature}).
The vast majority of the dwarf, low-mass stars in the metallicity range between [Fe/H]$\sim$--3 and $\sim$--1 dex 
share a similar abundance, $\rm A(Li)_{NLTE}$ $\sim$2.2-2.3 dex. At higher metallicities two effects operate, destroying the 
plateau and significantly increasing the star-to-star scatter. The first one is the presence in these stars of 
more massive convective envelopes, leading to more efficient surface A(Li) depletion \citep{melendez14}. 
The second effect is the occurrence of novae that produce fresh Li and contribute to the chemical 
enrichment of the Galaxy for [Fe/H]$>$--1 dex \citep[see e.g.][]{izzo15,romano21,izzo23}. 
Actually after the detection in RS Oph \citep{molaro23}
we know that also recurrent novae contribute to the Li production.
In this way, the interpretation of metal-poor Li-rich stars has the advantages to remove from the discussion 
the effects of the extra-dilution due to massive convective envelope and the effect of novae producing additional Li.

In principle, A(Li) of a Li-rich star (regardless of its origin) should follow 
the same evolutionary path of a Li-normal star, with a significant reduction at the FDU and the RGBb. 
For this reason, Li-rich stars should be discussed considering their evolutionary stage 
and the possible occurrence of the mixing episodes.

\begin{figure*}
\centering
\includegraphics[scale=0.8]{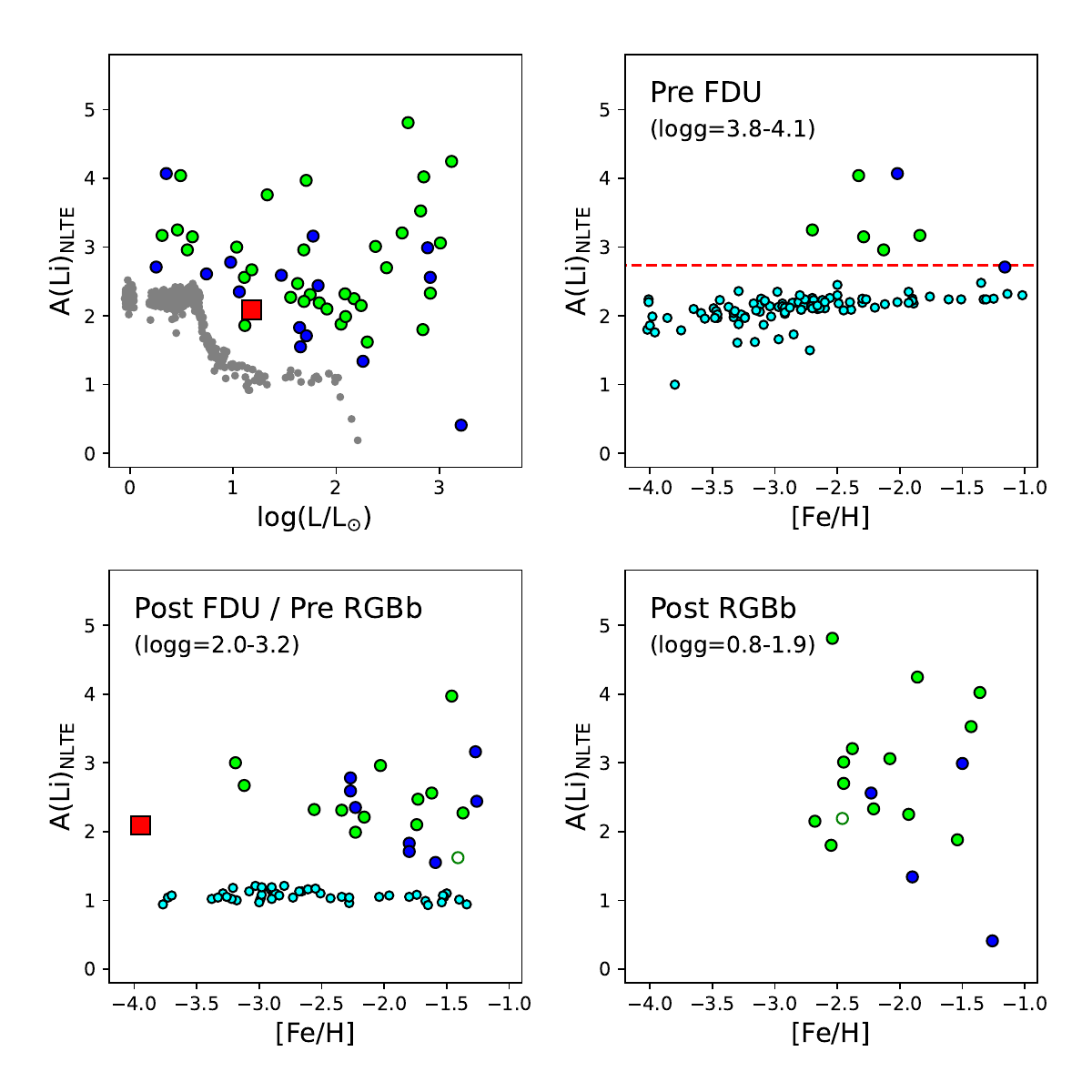}
\caption{Upper-left panel: 
behaviour of \ali\ as a function of \lum\ for all the metal-poor ([Fe/H]$<$--1 dex) Li-rich stars discovered so far.
Green circles are Milky Way field stars, blue circles are globular clusters stars and the red squares 
marks \he . The stars of the metal-poor ([Fe/H]$\sim$--2.0 dex) globular cluster NGC~6397 \citep[grey circles,][]{lind09} are shown 
as a reference (and to identify the luminosity of the FDU and RGBb).
Upper-right panel: run of \ali\ as a function of [Fe/H] for the Li-rich stars located before the FDU.
Red dashed line indicates the WMAP/SBNN A(Li) \citep{coc17}. The cyan circles are Li-normal Milky Way field stars 
\citep{ryan99,lucatello03,ivans05,asplund06,sivarani06,thompson08,sbordone10,masseron12,ito13,hansen14,bonifacio15,li15,placco16,matsuno17,aguado18,bonifacio18}. 
Lower-left panel: run of \ali\ as a function of [Fe/H] for the Li-rich stars located after the completion of FDU 
and before the RGBb. Open circle indicates the star 2MASS J19524490-6008132	with an uncertain attribution to this group.
The cyan circles are Li-normal Milky Way field stars \citep{mucciarelli22}.
Lower-right panel: run of \ali\ as a function of [Fe/H] for the Li-rich stars located after the RGBb. 
Open circle indicates the star GSC 03797-00204 with an uncertain attribution to this group.
}
\label{lit3}
\end{figure*}

The upper-left panel of Fig.~\ref{lit3} shows the run of \ali\ as a function of \lum\ for 
\he\ and all the other metal-poor Li-rich stars discovered so far. 
Despite a significant star-to-star scatter, it is possible to recognise 
some sequences where A(Li) decreases with increasing \lum . Some of the field Li-rich stars 
draw a clear sequence starting from A(Li)$\sim$3.2 dex at \lum\ $\sim$+0.5 down to 
A(Li)$\sim$2.2 dex at \lum\ $\sim$+2. The GC Li-rich stars in the same luminosity range seem to draw a parallel 
sequence but shifted by 0.4 dex toward lower A(Li). Finally, some field and GC stars define a  
super Li-rich sequence, with values higher by 1 dex than the other Li-rich of similar \lum .
Two sequences of A(Li) as a function of \lum\ for the GC Li-rich stars have been already proposed by \citet{sanna20}.
Note that the Li-rich stars show a decrease of A(Li) less steep than that expected by the FDU.

The other panels of Fig.~\ref{lit3} show the behaviour of \ali\ as function of [Fe/H] for the stars 
grouped according to their evolutionary stages. In particular, we consider stars located before the FDU 
(corresponding to the Spite Plateau for Li-normal stars), 
after the FDU and before the RGBb (corresponding to the lower RGB Plateau for Li-normal stars) and 
after the RGBb (stars experienced the extra-mixing episode at the luminosity level of the RGBb).
One only Li-rich star, namely 132 in the GC M30, is clearly located during the FDU (see Fig.~\ref{lit1} and Table~\ref{tab_gc}) and excluded from this discussion. 
Also, two field stars (namely GSC 03797-00204 and 2MASS J19524490-6008132) are close the RGBb and we propose an attempt 
at classification (they are marked as empty symbols 
in Fig.~\ref{lit3}, see Table~\ref{tab_field}). Finally, for two stars 
(namely SDSS J143207.14+081406.1 and 2MASS J04315411-0632100) the attribution is too uncertain 
due to their anomalous position in the Hertzsprung-Russell diagram (see Fig.~\ref{lit2}).

Seven Li-rich stars (two of them members of GCs) are located before the occurrence of the FDU. 
Six of them have \ali\ exceeding the primordial value obtained from the standard Big Bang nucleosynthesis model 
and the Planck/WMAP measures of the baryon density.
Among them, four stars have \ali $\sim$3-3.2 dex (about 1 dex higher than the Spite Plateau), and other two stars have \ali $\sim$4 dex (about 2 dex higher than the Spite Plateau).
Only the Li-rich star in M4 \citep{monaco12} has a value compatible with the primordial value.

Twenty-one stars are located between the completion of the FDU and before the RGBb. 
They show a large A(Li) scatter and a typical value around +2.4 dex, 1.4 dex higher than the abundances measured 
along the lower RGB Plateau. However, their average value is lower than that measured in the previous group, 
suggesting that a dilution due to the FDU occurred. 

Eighteen stars (4 of them members of GCs) are located after the RGBb. 
In this group we found the Li-rich stars with the lowest \citep{smith99} and the highest 
\citep{kow22} \ali\ of the entire sample. The average value of these stars 
well matches that of the stars between the FDU and the RGBb.

For the stars between FDU and the RGBb, we corrected the measured \ali\ to take into 
account the effect of the Li dilution due to the FDU, following the method described 
in \citet{mucciarelli12}. 
In particular, for each star in this evolutionary stage, we consider the stellar model with the appropriate metallicity
that provides the amount of A(Li) dilution as a function of the stellar 
luminosity. The A(Li) dilution to be added to the measured \ali\ value is computed according to the 
\lum\ value of each star. This approach is restricted to stars experiencing the FDU but fainter than the RGBb, 
because the standard stellar evolution models that we adopted do not account for non canonical mixing processes.

\begin{figure}
\centering
\includegraphics[width=\hsize,clip=true]{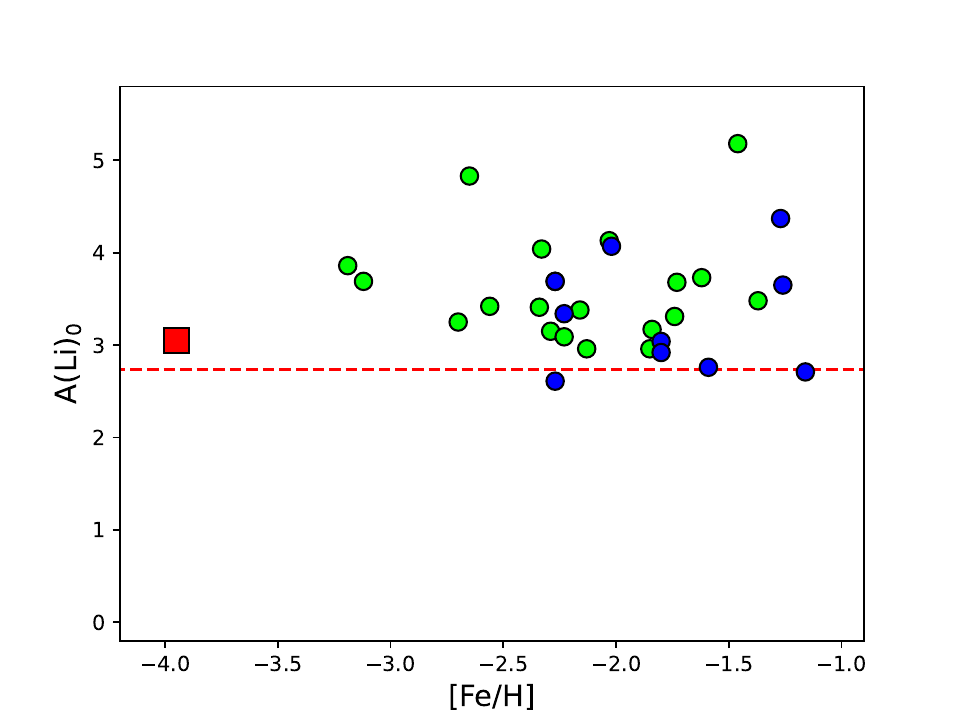}
\caption{Behaviour of the initial A(Li), accounting for the dilution effect due to the FDU, for \he\ and 
other metal-poor Li-rich stars fainter than the RGBb (same symbols of Fig.~\ref{lit2}).}
\label{lit4}
\end{figure}

Fig.~\ref{lit4} shows the run of the corrected A(Li) with [Fe/H] 
for these stars, together with the (uncorrected) abundances for stars before the FDU.
No evident trend between the initial A(Li) and [Fe/H] is found. The derived distribution 
exhibits a large star-to-star scatter and almost all the stars (assuming that they formed 
with a high A(Li)) have abundances higher than the cosmological value.
For the target star \he\ the predicted initial A(Li) should be +3.05 dex, $\sim$0.3 dex higher 
than the cosmological value. 
Only a few stars could have an initial A(Li) compatible with the cosmological value 
(and therefore explainable by invoking some preservation of the pristine lithium),
while for most of the Li-rich stars processes able to produce or enhance the surface lithium abundance 
should occur.

\subsection{\he : a Na-rich Li-rich metal-poor star}

\begin{figure}
\centering
\includegraphics[width=\hsize,clip=true]{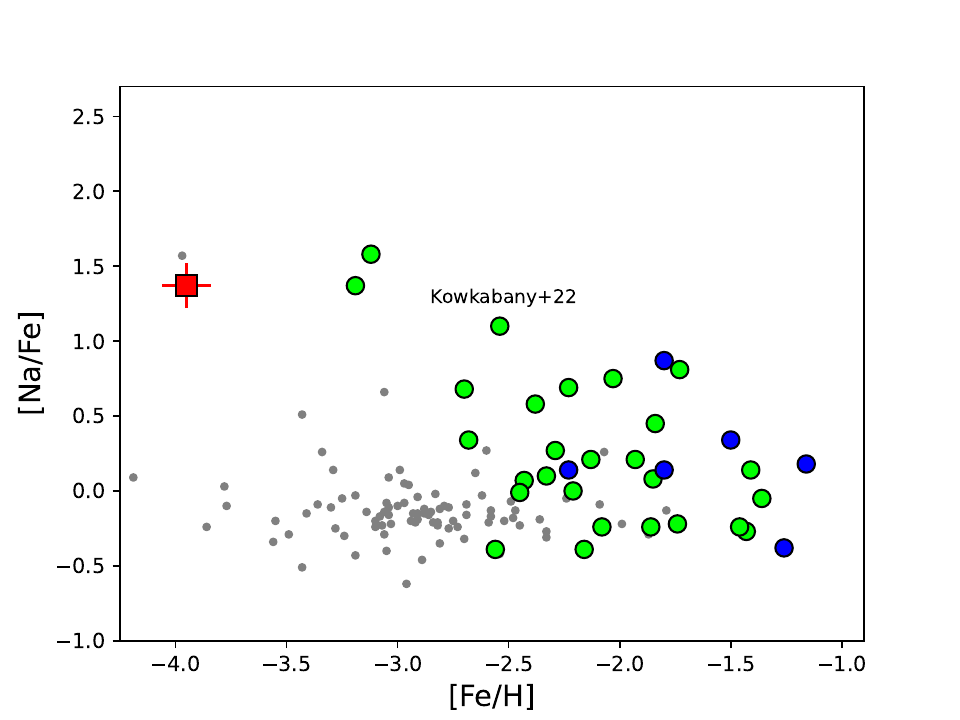}
\caption{Behaviour of [Na/Fe] as a function of [Fe/H] of the metal-poor Li-rich stars 
(same symbols of Fig.~\ref{lit2}) in comparison with the NLTE abundances for metal-poor 
Milky Way stars \citep[grey circles, from][]{andr07,lombardo22}. The [Na/Fe] abundance 
ratio of \he\ plotted here accounts for NLTE effects for Na abundance only. The position of the 
Li-rich star by \citet{kow22}, the most Li-rich discovered so far, is marked.}
\label{sod}
\end{figure}

Previous works about metal-poor Li-rich stars highlight significant over-abundances of [Na/Fe] 
in some of these stars \citep[see e.g.][]{kow22,sitnova23} but this chemical signature has not been properly discussed.
Na abundances are available for 23 field stars, shown in Fig.~\ref{sod} in comparison 
with the metal-poor Milky Way field stars \citep{andr07,lombardo22}.
We show also the 6 GC Li-rich stars with [Na/Fe] abundances even if the discussion of these stars is complicated 
by the self-enrichment processes occurring in the early epochs of the GC life and likely able to 
form new stars with excess of [Na/Fe] \citep[see e.g.][]{bastian}. In the following we refer only to 
the field Li-rich stars.

The [Na/Fe] distribution of the metal-poor Li-rich stars does not match 
that of the Milky Way field stars. For [Fe/H]$<$--2.0 dex, the field stars 
have a constant value of [Na/Fe]$\sim$--0.2 dex, while the Li-rich stars exhibit a significant 
star-to-star scatter in [Na/Fe], reaching very high values, up to [Na/Fe]$\sim$+1.6 dex.
In particular, the three Li-rich stars with [Fe/H]$<$--3.0 dex have [Na/Fe]$>$+1.3 dex and are the most 
Na-rich Li-rich stars: \he\ with [Na/Fe]=+1.37 dex, Gaia EDR3 883042050539140992 with [Na/Fe]=+1.37 dex \citep{li18} 
and Gaia EDR3 2604066644687553792 with [Na/Fe]=+1.58 dex \citep{roederer14}. In this comparison and 
in Fig.~\ref{sod} we used for \he\ [Na/Fe]=+1.37 obtained by adopting NLTE Na abundance and LTE Fe abundance, 
similar to the analyses of other Li-rich stars, where only the Na abundances are corrected for NLTE effects.

Even if only three Li-rich stars with [Fe/H]$<$--3 dex have been discovered so far, 
their extremely high [Na/Fe] abundance ratios 
could be a new characteristic feature of this class of rare objects not explored before. 
Sodium is produced both in massive stars during the hydrostatic C and Ne burning, and in AGB stars  
during the hot bottom burning phase. In particular, super-AGB stars, with initial masses larger than $\sim$6-7 $M_{\odot}$
should be able to produce large amount of both Li and Na, at  least for [Fe/H]$>$--2.5 dex 
\citep{ventura11,dantona12,doherty14}. 
In these stars Li is produced through the Cameron-Fowler mechanism and Na through the Ne-Na cycle.
The evidence that all the three Li-rich stars with [Fe/H]$<$--3 dex have an excess of [Na/Fe] could be 
an important hint to support the scenario of a mass transfer process occurring in binary systems where 
the companion was a massive star able to produce simultaneously Li and Na. Unfortunately, theoretical models 
for AGB stars at [Fe/H]=--4 dex are not available so far.

\section{Conclusions}
We revised the nature of the metal-poor star \he , demonstrating that it is a 
genuine Li-rich star, joining the limited class of the metal-poor Li-rich stars. 
Its very low metallicity demonstrates that at least one of the proposed mechanisms able to produce Li-rich stars 
should work down to [Fe/H]$\sim$--4 dex. 

However, we are not yet able to surely identify the process capable of producing the excess of lithium observed in this star 
and generally in other metal-poor stars. 
This is due both to our still partial understanding of the properties of these stars and to the lack of a sound mechanism 
able of triggering the Cameron-Fowler mechanism in many of these stars (especially those fainter than the RGBb). 
For the star \he , we can only speculate that the excess of lithium may be attributable to a mass transfer process 
in a binary system, despite not having strong evidence, except perhaps the very high [Na/Fe], supporting this hypothesis.
The extremely high Li and Na abundances could be compatible with a mass transfer from a companion in the stellar range of 
6-8 $M_{\odot}$, while the lack of RV variations from the available three epochs and the RUWE Gaia parameter close to the unity do not 
support the binary nature of the star (but neither rule out a long-period binary system). 
On the other hand, \he\ does not exhibit enhancement of neutron-capture elements 
(Sr and Ba) usually associated to the mass transfer process from AGB stars. However, it is important to bear in mind that 
AGB stars with 2-4 $M_{\odot}$ produce a large amount of neutron capture elements while 
stars with 6-8 $M_{\odot}$ are extremely less efficient to produce these elements \citep[see e.g.][]{fishlock14,shingles15}. 
Therefore, all the chemical evidence collected so far for \he\ agree with this scenario.

Detailed chemical abundances of the main groups of elements are limited to a few stars, 
in particular elements tracer of mass transfer from AGB stars (i.e., CNO, $^{12}$C/$^{13}$C and neutron capture processes) 
are simultaneously available only for 9 Li-rich stars.
Another missing piece of evidence are dedicated surveys of RV in order to monitor possible 
variability and establish the binary nature of some of these stars. 
In the same way, systematic studies of other diagnostics of interactions (chromospheric activity, stellar rotation...) 
are still lacking.
Metal-poor Li-rich stars are therefore an unexplored field of research that still deserves deep investigations.

\begin{landscape}
\begin{table*}\tiny
\caption{Main information about the Li-rich stars discovered so far in the Milky Way field. 
The stars are in order of the Gaia EDR3 identification number (we report also the alternative identification number or name used 
in the literature).
Stellar parameters and \ali\ are those described in Section~\ref{par}.
[Fe/H] and [Na/Fe] are those listed in the corresponding papers. The evolutionary stage of each target is obtained as described in  
Section~\ref{literature}: (1) pre-FDU, (2) FDU, (3) post-FDU / pre-RGBb, (4) RGBb.
References: (1) \citet{li18}, (2) \citet{martell13} , (3) \citet{sitnova23} , (4) \citet{roederer14}, (5) \citet{ruchti11}, 
(6)  \citet{kow22}, (7)  \citet{roederer08}, (8) this study, (9) \citet{susmitha24}.}
\label{tab_field}
\begin{tabular}{ccc cc cc ccc}
\hline\hline
${\rm Gaia~EDR3}$  & ${\rm ID}$   & \teff\ & \gr\  &    ${\rm [Fe/H]}$ &  \lum\ & \ali\  &  [Na/Fe] &  stage &  ${\rm REF}$ \\
\hline 
       &       &   (K)   &   (dex)   &   (dex)  &  (dex)    &  (dex)  & (dex)  & &   \\
\hline	
   29184921251610880   & LAMOST J030209.33+135656.3	 &    5118  &	2.22 &   --1.74  &    1.91$\pm$0.03   &	2.10   & --0.22  & 3 & 1	    \\     
  215823657109094528   & LAMOST J055408.54+523559.0	 &    5308  &	2.51 &   --2.03  &    1.69$\pm$0.04   &	2.95   &   0.75  & 3 & 1	   \\	  
  673258221256238464   & LAMOST J074102.07+213246.6	 &    6189  &	3.97 &   --2.33  &    0.49$\pm$0.02   &	4.05   &   0.10  & 1 & 1	  \\	 
  691015471484504704   & LAMOST J085208.07+262730.1	 &    5887  &   3.82 &	 --2.13  &    0.56$\pm$0.05   & 2.96   &   0.21  & 1 & 1	   \\	  
  696658714913856640   & SDSS J093627.44+293535.7	 &    5263  &   3.07 &	 --1.62  &    1.11$\pm$0.20   & 2.55   &    ---  & 3 & 2   \\  
  883042050539140992   & LAMOST J070542.30+255226.6	 &    5287  &   3.15 &	 --3.19  &    1.04$\pm$0.02   & 3.01   &   1.37  & 3 & 1    \\     
  932683488703778048   & LAMOST J075816.39+470343.3	 &    5897  &   4.06 &	 --1.84  &    0.31$\pm$0.09   & 3.16   &   0.45  & 1 & 1	   \\	  
 1005288880051257088   & LAMOST J062647.91+603254.0	 &    5962  &	3.79 &   --2.29  &    0.61$\pm$0.02   &	3.16   &   0.27  & 1 & 1	  \\	 
 1030551568447702400   & GSC 03797-00204		 &    4760  &	2.16 &   --2.46  &    1.84$\pm$0.08   &	2.19   &    ---  & 4(?)       & 2   \\  
 1162478360892154112   & 2MASS J15221187+0655551	 &    5167  &	2.40 &   --2.34  &    1.75$\pm$0.11   &	2.31   &    ---  & 3 & 2   \\  
 1172619465872571776   & SDSS J143207.14+081406.1	 &    4661  &	2.62 &   --2.65  &    1.35$\pm$0.45   &	3.75   &    ---  & N/A   & 2   \\  
 1181606805198585472   & LAMOST J145500.04+125106.2	 &    4830  &	1.78 &   --2.68  &    2.25$\pm$0.06   &	2.15   &   0.34  & 4  & 1	   \\	  
 1522817217854702592   & LAMOST J131457.78+374110.7	 &    5992  &	3.94 &   --2.70  &    0.46$\pm$0.02   &	3.26   &   0.68  & 1  & 1	   \\	  
 1800137243204885120   & LAMOST J214610.13+273200.8	 &    5187  &	2.52 &   --1.73  &    1.63$\pm$0.11   &	2.47   &   0.81  & 3 & 1	   \\	  
 1918529631627603072   & ---				 &    4649  &	1.58 &   --2.43  &    2.38$\pm$0.07   &	3.01   &   0.07  & 4	      & 3    \\
 2604066644687553792   & BPS CS22893-010		 &    5374  &	3.03 &   --3.12  &    1.18$\pm$0.07   &	2.68   & --0.05  & 3 & 4  \\	
 3199294789168393600   & 2MASS J04315411-0632100	 &    5662  &	3.19 &   --1.85  &    1.12$\pm$0.01   &	1.86   &   0.08  & N/A  	      & 5   \\  
 3210839729979320064   & 2MASS J05241392-0336543	 &    4240  &	1.10 &   --2.54  &    2.70$\pm$0.06   &	4.79   &   1.10  & 4	      & 6	\\     
 3360259919923274496   & LAMOST J071422.66+160042.5	 &    5116  &	2.44 &   --2.16  &    1.69$\pm$0.02   &	2.21   & --0.39  & 3 & 1	   \\	  
 3665949720385847040   & LAMOST J141412.27+001618.7	 &    5033  &	2.01 &   --2.56  &    2.09$\pm$0.07   &	2.32   & --0.39  & 3 & 1	   \\	  
 3687632330203892352   & GSC 04958-01069		 &    4580  &	1.87 &   --1.54  &    2.07$\pm$0.07   &	1.88   &    ---  & 4	      & 2   \\  
 3926388527100945920   & UCAC2 37720962		         &    5196  &	2.06 &   --2.23  &    2.09$\pm$0.14   &	1.99   &   0.69  & 3 & 7  \\	
 4425538154386975616   & 2MASS J16070923+0447126	 &    5105  &	2.56 &   --1.37  &    1.56$\pm$0.31   &	2.28   &    ---  & 3 & 2    \\   
 4903905598859396480   & HE0057-5959			 &    5420  &	3.05 &   --3.98  &    1.18$\pm$0.14   &	2.09   &    ---  & 3 & 8    \\ 		 
 5441530913278761088   & UCAC4 253-045343                &    4285  &   1.00 &   --1.43  &    2.82$\pm$0.07   & 3.53   & --0.27  & 4 & 9    \\
 5668925276702015872   & 2MASS J10122548-2030068	 &    4451  &	1.05 &   --2.55  &    2.84$\pm$0.36   &	1.77   &    ---  & 4 & 5 \\	 
 6142572036722901504   & TYC 7262-250-1  		 &    4179  & 	0.92 &	 --1.36  &    2.85$\pm$0.02   & 4.02   & --0.05  & 4 & 9   \\
 6221353316163376768   & UCAC4 308-077592		 &    4086  & 	0.62 &	 --1.86  &    3.12$\pm$0.13   & 4.25   & --0.24  & 4 & 9   \\
 6298610530751953920   & 2MASS J14254628-1546301	 &    4287  &	0.81 &   --2.08  &    3.01$\pm$0.06   &	2.96   & --0.24  & 4 & 5    \\ 
 6381456605897426560   & UCAC4 099-098976		 &    5291  & 	2.48 &	 --1.46  &    1.71$\pm$0.01   & 3.97   & --0.24  & 3 & 9   \\
 6425395014689839744   & TYC 9112-00430-1		 &    4394  &	0.95 &   --2.21  &    2.91$\pm$0.05   &	2.27   &   0.00  & 4 & 5    \\ 
 6443920273789215872   & 2MASS J19524490-6008132	 &    5029  &	1.79 &   --1.41  &    2.30$\pm$0.02   &	1.62   &   0.14  & 3(?) & 5	\\	 
 6529464133855343872   & TYC 8448-00121-1		 &    4731  &	1.50 &   --2.45  &    2.49$\pm$0.06   &	2.71   & --0.01  & 4 & 5    \\  
 6668577437377879424   & UCAC4 212-183136		 &    4308  & 	1.19 &	 --2.38  &    2.64$\pm$0.07   & 3.21   &  +0.58  & 3 & 9  \\
 6816797308517921792   & TYC 6953-00510-1		 &    4836  &	1.85 &   --1.93  &    2.18$\pm$0.03   &	2.25   &   0.21  & 4 & 5   \\

\hline
\end{tabular}   
\end{table*}
\end{landscape}

\begin{landscape}
\begin{table*}\tiny
\caption{Main information about the Li-rich stars discovered so far in Galactic globular clusters. 
The clusters are in order of right ascension. We list for each target the Gaia ED3 identification number 
and the alternative identification number or name used in the literature.
Stellar parameters and \ali\ are those described in Section~\ref{par}.
[Fe/H] and [Na/Fe] are those listed in the corresponding papers. The evolutionary stage of each target 
is obtained as described in Section~\ref{literature}: (1) pre-FDU, (2) FDU, (3) post-FDU / pre-RGBb, (4) RGBb.
References: (1) \citet{smith99},
(2) \citet{dorazi15},
(3) \citet{sanna20},
(4) \citet{agu22},
(5) \citet{ruchti11},
(6) \citet{kirby16},
(7) \citet{mucciarelli19},
(8) \citet{kraft99},
(9) \citet{monaco12},
(10) \citet{koch11},
(11) \citet{gruyters16}.}
\label{tab_gc}
\begin{tabular}{cc cc cc cc cc c}
\hline\hline
{\rm CLUSTER} & ${\rm Gaia~EDR3}$  & ${\rm ID}$   & \teff\ & \gr\  &    ${\rm [Fe/H]}$ &  \lum\ & \ali\  & [Na/Fe] &  stage &  ${\rm REF}$ \\
\hline 
     &  &       &   (K)   &   (dex)   &   (dex)  &  (dex)    &  (dex)   & (dex) &  &   \\
\hline\hline
N362   & 4690838559145577728  &  V2				 &  3651   &  0.33  & -1.26  &  3.21 &     0.41  &  --0.38  &	 4	       &      1  \\
N362   & 4690838726645395840  &  15370  			 &  5004   &  2.27  & -1.26  &  1.82 &     2.44  &     ---  &	 3    &      2  \\ 
N1261  & 4733703810220702976  &  GES J03115070-5514001  	 &  4984   &  2.31  & -1.27  &  1.78 &     3.16  &     ---  &	 3    &      3  \\ 
N3201  & 5413582186704791168  &  97812  			 &  5001   &  2.44  & -1.59  &  1.65 &     1.55  &     ---  &	 3    &      4  \\
M68    & 3496374958317915648  &  Stet-M68-S232  		 &  4488   &  0.99  & -2.23  &  2.91 &     2.56  &    0.14  &	 4	       &      5,6 \\ 
M68    & 3496399735985613312  &  Stet-M68-S534  		 &  5397   &  3.16  & -2.23  &  1.06 &     2.35  &     ---  &	 3    &      6   \\
N5053  & 3938494188778278272  &  N5053-S79			 &  5277   &  2.71  & -2.27  &  1.47 &     2.59  &     ---  &	 3    &      6   \\
OCen   & 6083508196239683968  &  25664  			 &  5061   &  2.46  & -1.80  &  1.65 &     1.83  &    0.87  &	 3    &      7  \\ 
OCen   & 6083703428274945280  &  126107 			 &  5059   &  2.40  & -1.80  &  1.71 &     1.71  &    0.14  &	 3    &      7  \\ 
M3     & 1454783694639968384  &  M3-IV101			 &  4283   &  0.93  & -1.50  &  2.89 &     2.99  &    0.34  &	 4	       &      5,8 \\ 
N5897  & 6252666548333610624  &  Tes01-WF4-703  		 &  4804   &  1.76  & -1.90  &  2.26 &     1.34  &     ---  &	 4	       &      6  \\ 
M4     & 6045462826174755328  &  37934				 &  5914   &  4.13  & -1.16  &  0.25 &     2.71  &    0.18  &	 2	       &      9   \\ 
N6397  & 5921744197274599040  &  6282				 &  6241   &  4.12  & -2.02  &  0.36 &     4.07  &     ---  &	 2	       &      10	  \\ 
M30    & 6816582040462535552  &  132				 &  5598   &  3.54  & -2.27  &  0.74 &     2.61  &     ---  &	 1		       &      6,11  \\ 
M30    & 6816574859276974336  &  7229				 &  5409   &  3.25  & -2.27  &  0.98 &     2.78  &     ---  &	 3    &      6   \\

\hline
\end{tabular}   
\end{table*}
\end{landscape}

\begin{acknowledgements}
We thanks the anonymous referee for the useful and constructive suggestions.
A.M. acknowledges support from the project "LEGO – Reconstructing the building blocks of the Galaxy by chemical tagging" 
(P.I. A. Mucciarelli). granted by the Italian MUR through contract PRIN
2022LLP8TK\_001. 
M.M. acknowledges support from the ERC Consolidator
Grant funding scheme (project ASTEROCHRONOMETRY, https://www.
asterochronometry.eu, G.A. n. 772293).
 This work has made use of
data from the European Space Agency (ESA) mission Gaia (https://www.
cosmos.esa.int/gaia), processed by the Gaia Data Processing and Anal-
ysis Consortium (DPAC, https://www.cosmos.esa.int/web/gaia/dpac/
consortium). Funding for the DPAC has been provided by national institutions,
in particular the institutions participating in the Gaia Multilateral Agreement.
This research has made use of the SIMBAD database, operated at CDS, Strasbourg, France.

\end{acknowledgements}

\begin{appendix}

\section{Asteroseismic data}
 
\begin{figure}[h]
\centering
\includegraphics[scale=0.48]{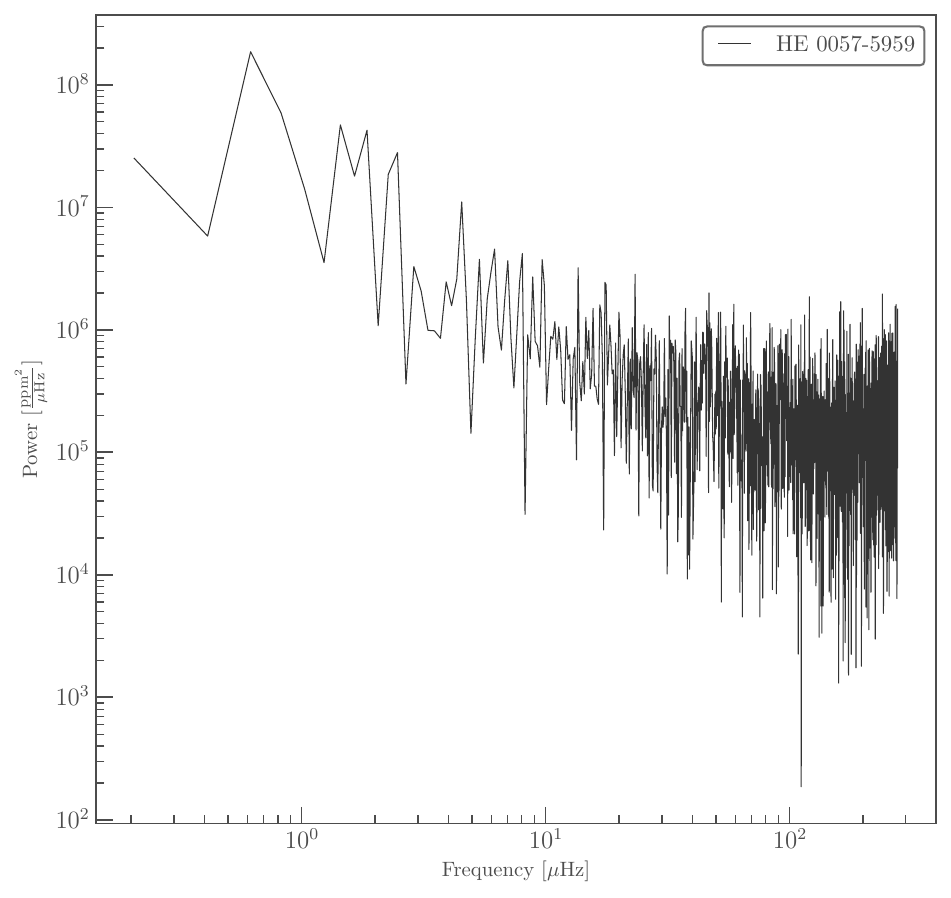}
\caption{Power spectral density for \he\ based on TESS observations during Sector 1 and 2. 
The light curve from which we obtain the power spectral density is provided by the TESS Asteroseismic Science Operations Center 
\citep[TASOC;][]{Handberg2021, Lund2021}. We find no evidence of solar-like oscillations.}
\label{aster}
\end{figure}

Here we discuss asteroseismic observations of metal-poor Li-rich stars (Table~\ref{tab_field}). 
Our sample of stars has been observed by TESS \citep[Transiting Exoplanet Survey Satellite;][]{ricker15},
but only 16 stars have available light curves in the Mikulski archive for space telescopes
(MAST\footnote{\url{https://archive.stsci.edu}}). We apply a Lomb–Scargle transform 
\citep[][]{Lomb1976,Scargle1982} to these lightcurves by means of the Python package 
{\scshape Lightkurve v2.4.1}\footnote{\citet{2018ascl.soft12013L}; \url{https://github.com/lightkurve/lightkurve}}. 
We do not find evidence of solar-like oscillations in any of these stars even with lightcurves based 
on the MIT quick-look pipeline \citep[QLP;][]{Huang2020}, or the TESS data for asteroseismology light curves pipeline 
\citep[TASOC; ][]{Handberg2021,Lund2021}. 
Fig.~\ref{aster} shows as example of the power spectral density for \he\ based on TESS observations.
Indeed, we could not find any of these stars in catalogues of solar-like oscillators observed with TESS \citep{Hon2021,Mackereth2021,Hatt2023}, 
and a similar result is shown in \citet{kow22} for the star Gaia DR3 3210839729979320064. 
Finally, these results suggest that more than 130 days (corresponding to longest observations made in our sample, 
that is those for Gaia DR3 3360259919923274496) are needed to observe solar-like oscillations in such stars.

\section{Information about the measured atomic lines}
\label{llist}

Table~\ref{llist2} lists the main atomic data (wavelength, oscillator strength, 
excitation potential and ion) for all the measured transitions used in the analysis.

\begin{table*}
\caption{}            
\label{llist2}     
\centering                          %
\begin{tabular}{c  c c  c }       
\hline\hline              
{\rm Wavelength}  & {\rm log~gf} & $\chi$ &    {\rm Ion}  \\
\hline
(\AA ) & (dex) & (eV) &      \\
\hline 
  6707.***  &  --0.002 &  0.00  &   {\rm Li~I}   \\
  5889.9**  &   0.108 &   0.00  &   {\rm Na~I}    \\
  5895.9**  &  --0.194 &   0.00  &  {\rm Na~I}      \\ 
  3838.292  &   0.397 &   2.72  &   {\rm Mg~I}       \\ 
  5167.321  &  --0.870 &   2.71  &   {\rm Mg~I}       \\ 
  5172.684  &  --0.393 &   2.71  &   {\rm Mg~I}      \\ 
  5183.604  &  --0.167 &   2.72  &   {\rm Mg~I}       \\ 
  3944.***  &  --0.635 &   0.00  &   {\rm Al~I}       \\ 
  3961.5**  &  --0.333 &   0.01  &   {\rm Al~I}      \\ 
  3905.523  &  --1.041 &   1.91  &   {\rm Si~I}     \\ 
  4226.728  &   0.244 &   0.00  &   {\rm Ca~I}      \\ 
  4300.043  &  --0.460 &   1.18  &   {\rm Ti~II}      \\ 
  4563.758  &  --0.795 &   1.22  &   {\rm Ti~II}       \\ 
  4571.971  &  --0.310 &   1.57  &   {\rm Ti~II}      \\ 
  3820.425  &   0.119 &   0.86  &   {\rm Fe~I}        \\
  3840.437  &  --0.506 &   0.99  &   {\rm Fe~I}      \\
  3841.048  &  --0.045 &   1.61  &   {\rm Fe~I}        \\
  3849.966  &  --0.871 &   1.01  &   {\rm Fe~I}       \\ 
  3878.018  &  --0.914 &   0.96  &   {\rm Fe~I}        \\
  3878.573  &  --1.379 &   0.09  &   {\rm Fe~I}      \\ 
  3886.282  &  --1.076 &   0.05  &   {\rm Fe~I}       \\ 
  3895.656  &  --1.670 &   0.11  &   {\rm Fe~I}        \\ 
  3902.945  &  --0.466 &   1.56  &   {\rm Fe~I}       \\ 
  3906.479  &  --2.243 &   0.11  &   {\rm Fe~I}       \\ 
  4005.241  &  --0.610 &   1.56  &   {\rm Fe~I}       \\
  4202.029  &  --0.708 &   1.48  &   {\rm Fe~I}     \\
  4250.786  &  --0.714 &   1.56  &   {\rm Fe~I}        \\
  4260.474  &	0.077 &   2.40  &    {\rm Fe~I}        \\
  4271.760  &  --0.164 &   1.48  &   {\rm Fe~I}     \\
  4325.762  &	0.006 &   1.61  &    {\rm Fe~I}        \\
  5227.188  &  --1.228 &   1.56  &   {\rm Fe~I}        \\
  5269.537  &  --1.321 &   0.86  &   {\rm Fe~I}         \\
  5270.356  &  --1.339 &   1.61  &   {\rm Fe~I}         \\
  5371.489  &  --1.645 &   0.96  &   {\rm Fe~I}        \\
  5397.127  &  --1.993 &   0.91  &   {\rm Fe~I}        \\
  5405.774  &  --1.844 &   0.99  &   {\rm Fe~I}        \\
  5446.916  &  --1.914 &   0.99  &   {\rm Fe~I}       \\
  4215.519  &  --0.173 &   0.00  &   {\rm Sr~II}      \\
  4554.0**  &	 0.140 &   0.00  &   {\rm Ba~II}     \\
 
\hline                                   
\end{tabular}
\end{table*}

\section{Comparison with previous chemical analyses}
\label{compar}

\begin{figure}[h]
\centering
\includegraphics[scale=0.6]{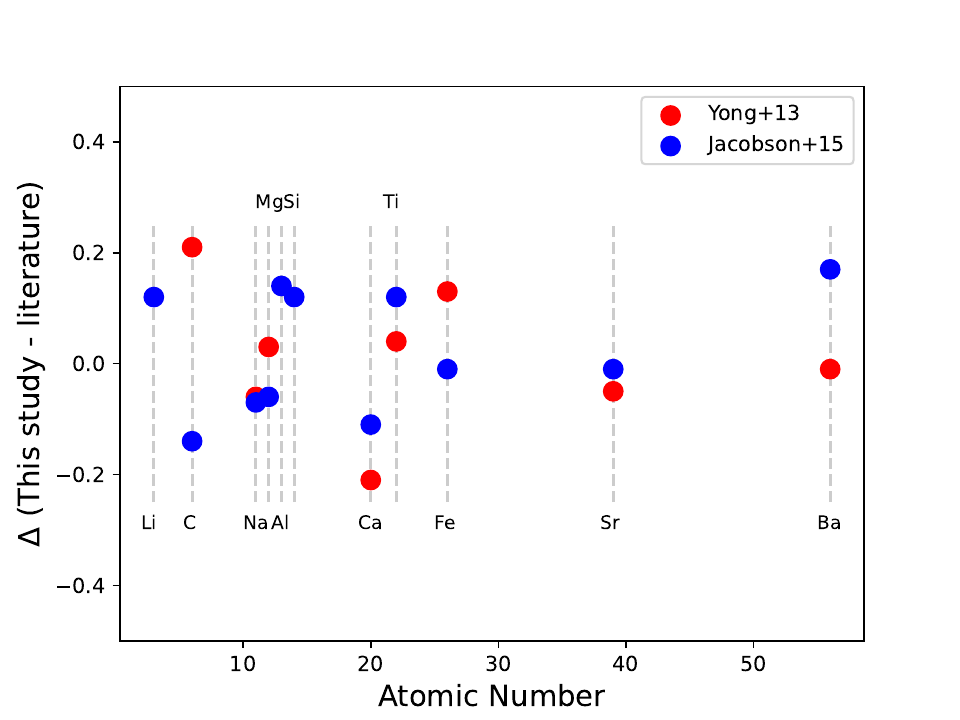}
\caption{Difference of our abundances with respect to those by \citet{yong13} and \citet{jacobson15}, 
red and blue points respectively, as a function of the atomic number. We consider our LTE abundances but for 
Li and Na that are provided by \citet{yong13} and \citet{jacobson15} corrected for NLTE effects.}
\label{compab}
\end{figure}

The chemical composition of \he\ has been already investigated by \citet{yong13} and \citet{jacobson15} 
both analysing MIKE spectra. Fig.~\ref{compab} shows the comparison between our analysis and those by 
\citet{yong13} and \citet{jacobson15} for the elements in common among the three studies. We consider 
our LTE abundances but for Li and Na because these two studies provided NLTE abundances only for these 
two elements. Our analysis well agrees with the previous ones, with differences that do not exceed 
$\pm$0.2 dex. The largest differences are for the [C/Fe] and [Ca/Fe] by \citet{yong13}, 0.21 dex lower and 
higher than our values, respectively. The differences with respect to these two studies can be ascribable 
to several differences in the chemical analyses, in particular, both the works adopted solar abundances by \citet{asplund09}, \citet{jacobson15} used MARCS model atmospheres, while \citet{yong13} adopted ATLAS9 model atmospheres like our analysis, and \citet{yong13} adopted a \teff\ value $\sim$200 K cooler than our one 
(see Section~\ref{atmpar}).

\section{Contamination of the Na D lines by interstellar features}

\begin{figure}[h]
\centering
\includegraphics[scale=0.6]{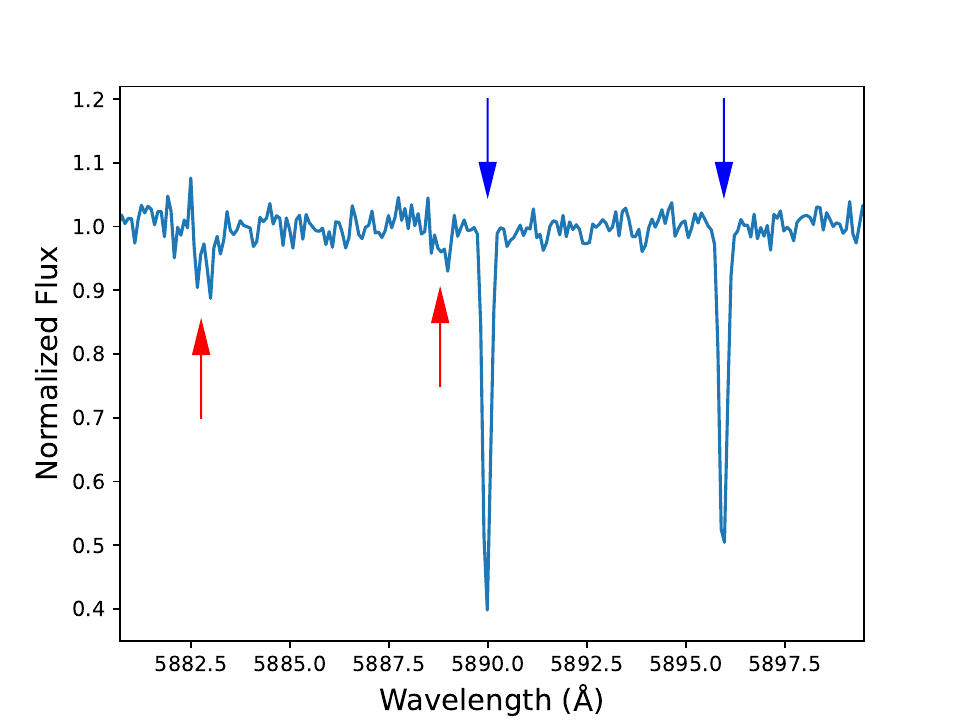}
\caption{Spectral region of the MIKE spectrum of \he\ with marked the photospheric and 
interstellar Na D lines (blue and red arrows, respectively).}
\label{naspec}
\end{figure}

The photospheric Na D lines at 5889.9 and 5895.9 \AA\ can be contaminated by the same transitions 
arising from the interstellar medium along the line of sight. We checked that the Na D lines in the 
spectrum of \he\ are not contaminated by interstellar features (see Fig.~\ref{naspec}) because of the large 
RV of the star. Also the interstellar lines are very weak because of the low colour excess of \he\ .

\end{appendix}

\end{document}